\documentstyle[aps,prl,preprint]{revtex}
\tightenlines
\begin{document}
% \draft command makes pacs numbers print
\draft
\title{Equilibrium Configurations and Energetics of Point Defects in Two-Dimensional Colloidal Crystals}
% repeat the \author\address pair as needed
\author{Alexandros Pertsinidis\cite{coresp} and X. S. Ling}
\address{Department of Physics, Brown University, Providence, Rhode Island 02912}

\date{\today}
\maketitle
\begin{abstract}
% insert abstract here
We demonstrate a novel method of introducing point defects (mono and di-vacancies) in a confined mono-layer colloidal crystal by manipulating individual particles with optical tweezers. Digital video microscopy is used to study defect dynamics in real space and time. We analyze the topological arrangements of the particles in the defect core and establish their connection with the energetics of the system. It is found that thermal fluctuations excite point defects into \textit{dislocation multipole} configurations. We extract the dislocation pair potential at near field, where cores overlap and linear elasticity is not applicable.
\end{abstract}
% insert suggested PACS numbers in braces on next line
\pacs{PACS number(s): 82.70.Dd, 05.40.Jc, 61.72.Ji, 61.72.Bb, 61.72.Ff}
% 82.70.Dd Colloids
% 05.40.Jc Brownian Motion
% 61.72.Ji Point Defects
% 61.72.Bb Theories & models of crystal defects
% 61.72.Ff Direct observation of dislocations and other crystal defects

% body of paper here
%\section{Introduction}
Colloidal crystals\cite{colloid_crystals}, ordered self-assembled structures of (sub)micron spheres, provide a model system for the study of basic problems in condensed matter physics. In particular, confined colloidal crystals have been used to explore statistical physics in two dimensions\cite{2D_colloid,2D_melting}. Although colloids in 2D have been studied for the last 20 years, several areas still remain relatively unexplored. The simplest structural defects in a crystal, namely vacancies and interstitials, fall in this category. Point defects are of considerable interest, since they play a dominant role in real materials' properties. Also, they are predicted to proliferate in more \textit{exotic} systems, such as the ground state of a Wigner crystal\cite{Wigner_crystal} or the supersolid phase\cite{supersolid} of the Abrikosov lattice in type II superconductors.

Colloidal crystals also offer new approaches in synthesizing materials with novel properties and applications like photonic-band-gap materials\cite{TiO2_opal}, optical switches\cite{Bragg_switch}, and chemical sensors\cite{chem_sens}. Understanding the behavior of defects in colloidal crystals, and inventing techniques for manipulating their dynamics can have an immediate impact on the fabrication of nanostructured materials.

In this work we demonstrate the use of optical tweezers\cite{tweezers}, a powerful tool that has promoted research in a wide range of fields, to artificially introduce isolated point defects in otherwise structurally perfect two-dimensional colloidal crystals. For the first time, such defects are studied by video microscopy in real space and time, providing new insight into their microscopic dynamics. In this letter the energetics of the defects are discussed. Results on the diffusion of the defects are presented in separate publication \cite{Part_B}.

%\section{Experimental details}
The colloidal crystals were prepared with a $\approx$1\%
Vol. aqueous suspension of 0.3 $\mu$m diameter negatively charged polystyrene-sulfate micro spheres (Duke Scientific, polydispersity $\simeq$1\%). The ionic strength of the suspension was reduced by ion exchange to a few $\mu$M. Under these conditions the screening length is of order a few hundred nm and the particles crystallize due to strong electrostatic interaction. Confining the suspension between two fused silica substrates separated by $\approx$2 $\mu$m suppresses the vertical motion of the spheres by negative charge that develops at the silica-water interface and creates a single layer colloidal crystal with a lattice constant $a\cong1.1 \mu$m\cite{c1}.

Trapping a particle with optical tweezers and dragging it from its lattice site creates isolated point defects in an otherwise perfect two-dimensional crystal\cite{Auxilliary}. The optical tweezers were formed by focusing a beam of an $\text{Ar}^{+}$ laser (Coherent INNOVA 90, $\lambda=$514nm) through the same objective (Zeiss Plan Neofluar, 100$\times$, oil-immersion, NA=1.3) used for imaging. The position of the beam was controlled by a set of mirrors mounted on an XY galvanometer scanner set  (Cambridge Technology). Point defects are created by dragging a particle away from its lattice point at about 50 $\mu$m/sec, making it an interstitial and leaving a vacant site behind\cite{di_creation}. To overcome the restoring force from the crystal and to be able to drag the particle faster than the relaxation time of the lattice, 100 mW of laser power ($\approx50$ mW going through the objective lens) was used.

The dynamics of such single, isolated point defects were studied in real time under 100$\times$ magnification and recorded on videotape (Sony SVO-9500MD recorder) using a monochrome CCD video-camera (Sony SSC-M370). The field of view covers ($50\times40$)$a^{2}$. Typically, we processed 60 frames/sec and tracked a single defect for 20-40 seconds, which is the time it takes to diffuse away from the field of view. Each individual frame was acquired on a PC and processed using a particle-tracking algorithm\cite{prtcl_track}. The raw data consists of the positions of the particles in every frame which are linked into trajectories in time.

%\section{Configurations of defects}
Real-space imaging facilitates detailed study of the structure of the defect core. The following questions naturally arise: what are the possible configurations in which a defect can exist? How can these be identified? What are their symmetries? Are there any geometrical-topological constraints that must be satisfied? In the case of strong interactions, $\kappa a\leq5.9$, where $\kappa^{-1}$ is the screening length, numerical studies\cite{supersolid,defect_configs} of mono-vacancies and interstitials have revealed that the system spontaneously deforms into configurations with lower symmetry than that of the lattice. This prediction is verified for our system, where $\kappa a\approx3$ (Fig. \ref{pic_configs}\textbf{A-D}). For di-vacancies, various stable configurations with well defined symmetry can also be identified (Fig. \ref{pic_configs}\textbf{E-H}).

The interactions in the system are short ranged (screened Coulomb, $V(r)\propto exp(-\kappa r)/r$), therefore the nearest neighbor bonds have geometrical as well as physical significance, since they identify the pairs of mutually interacting particles. The creation of a point defect involves breaking of bonds and creation of mis-coordinated particles in the core region. The configuration of the defects can be characterized by the arrangement of these broken bonds and mis-coordinated particles.
We analyzed the topological structure of the defects by performing a geometrical triangulation of the positions of the particles, finding the pairs of nearest neighbors, and measuring the coordination number $n_{i}$ for every particle. Our observations indicate that the following constraints are satisfied: (i)the mis-coordinated particles are never observed isolated but rather appear as \textit{distinguishable} pairs, triplets, etc. of nearest neighbors, (ii)the mean coordination number of every pair, triplet, etc. is always equal to six, (iii)we can define vectors $\vec{n}_{ij}$ starting from a particle i with $n_{i}<6$ and terminating to a nearest neighbor j with $n_{j}>6$. The number of vectors originating from i is $6-n_{i}$ and the number of vectors terminating on j is $n_{j}-6$. For every configuration, $\sum_{<ij>}^{}\vec{n}_{ij}\approx0$.

The finite temperature causes the system to vibrate around every local energy minimum $C$, exploring the volume of phase space $\Omega_{C}$ in which the energy is $O(k_{B}T)$ above the minimum energy $E_{C}$, and occasionally getting enough energy to jump to a nearby local minimum. 
The contribution to the entropy of every configuration is roughly $S_{C}=klog(\Omega_{C})$, and the different configurations occur with relative probabilities $P_{C}\propto exp(-F_{C}/k_{B}T)$, where $F_{C}=E_{C}-TS_{C}$ is the \textit{free} energy of a configuration.
As long as the system remains around a local energy minimum, the distortions of the lattice are elastic and the topological arrangement of the particles does not change.
Therefore, every energy minimum has not only a characteristic symmetry but also a certain topological configuration of the defect core. Using the topology rather than the symmetry as a criterion for identifying the different configurations $C$, we measure their relative occurence probabilities $P_{C}$.
The free energy differences, estimated from the measured $P_{C}$'s, are found to be of order 0.1-1 $k_{B}T$.

%\section{Point defects as fluctuating dislocation "multipoles"}
Point defects are "topologically neutral" since they have zero Burgers vector. However, one could think of creating a point defect by inserting extra rows of particles that do not terminate on each other. The extra inserted rows are edge dislocations whose Burgers vectors add up to zero. Therefore, a point defect contains embryonic pairs, triplets etc. of dislocations, in the same way that a dislocation contains an embryonic pair of a seven and a five-fold disclination \cite{defect_unbind}. Since the system is at finite temperature, one would expect the defect to fluctuate between these possible configurations of dislocations.

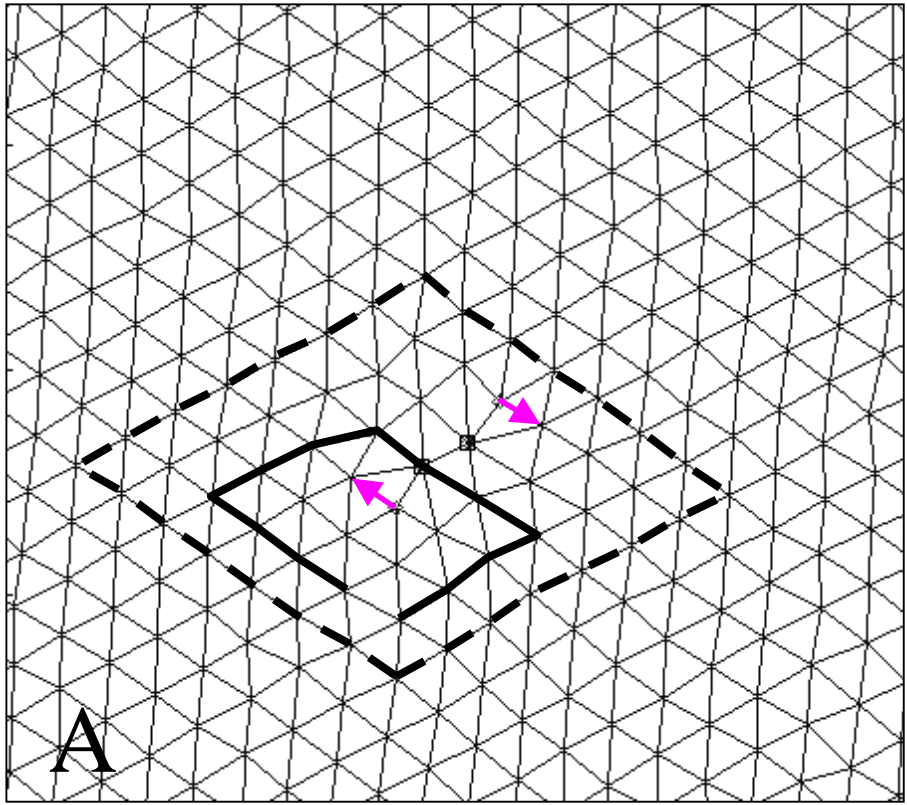
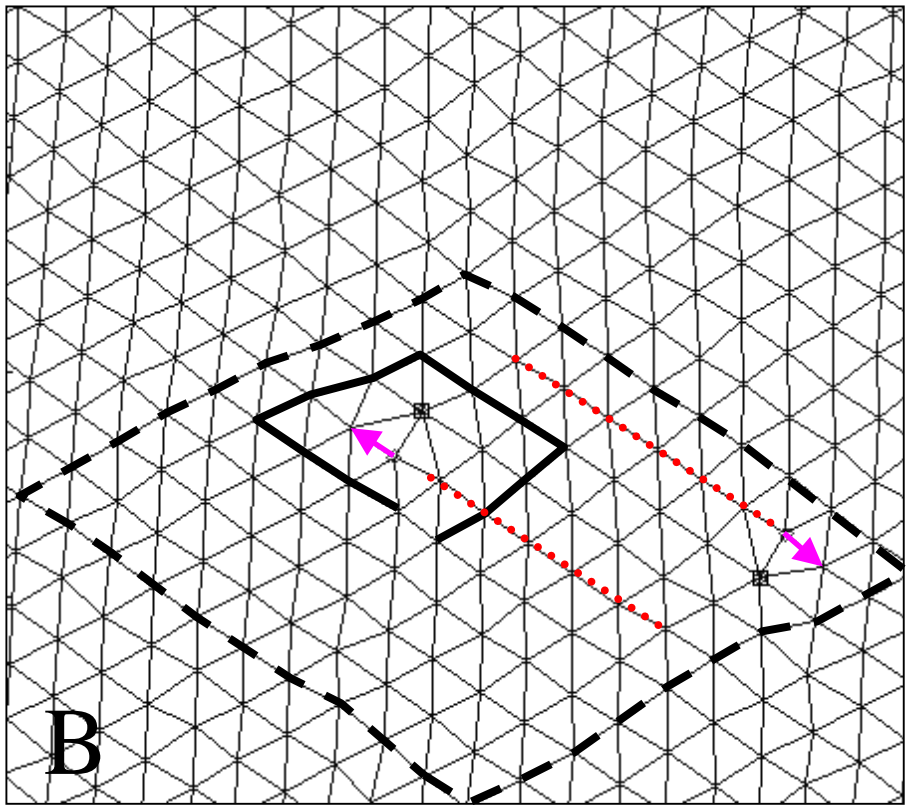
\begin{figure}
\caption{Point defects as dislocation multipoles: (\textbf{A}) well-defined di-vacancy configuration. A Burgers circuit (solid line) fails to close if it crosses the core of the defect. A Burgers circuit that surrounds the defect core without crossing it (dashed line) closes as expected. The arrows indicate the Burgers vectors of the "embryonic" dislocation dipole. (\textbf{B}) configuration of the same di-vacancy about 2 seconds later. The di-vacancy is not well defined any more and the system resembles a pair of dislocations with opposite Burgers vectors, whose gliding lines (dotted lines) are separated by $\approx3\vec{a}$. When those two dislocations come close to each other, the original di-vacancy is recovered.}
\label{disloc_multip}
\end{figure}

The signature of an isolated edge dislocation in a two-dimensional hexagonal lattice is a core with a pair of neighbor 5-fold and 7-fold coordinated particles. The extra row of particles terminates on the 5-fold particle. The Burgers vector $\vec{b}=\vec{a}$ is in 1-1 correspondence with the vector $\vec{n}_{ij}$ defined earlier in the text. $\vec{b}$ connects the 5-fold coordinated particle with its first nearest neighbor, moving counter-clockwise after the 7-fold neighbor (see Fig.\ref{disloc_multip}). Since $\vec{b}\leftrightarrow\vec{n}_{ij}$ and $\sum_{<ij>}^{}\vec{n}_{ij}\approx0$, one can view the various configurations of the defect as $\sum_{}^{}\vec{b}=0$ \textit{dislocation multipoles}.
Tracking the time evolution of the system, we see that the vast majority of the observed configurations correspond to a dislocation pair or a triplet, with infrequent appearance of higher order multipoles.
Most of the time the dislocations comprising the defect are closely bound in the configurations previously identified (Fig.\ref{pic_configs}). Occasionally however configurations are observed in which the dislocations appear to dissociate and recombine (Fig.\ref{disloc_multip}). This fits well with a picture in which one sees the dislocations as point particles interacting with an attractive potential. The bound configurations are minima of the energy and the system gets random kicks from the finite temperature background getting exited into states in which the dislocations can be separated.

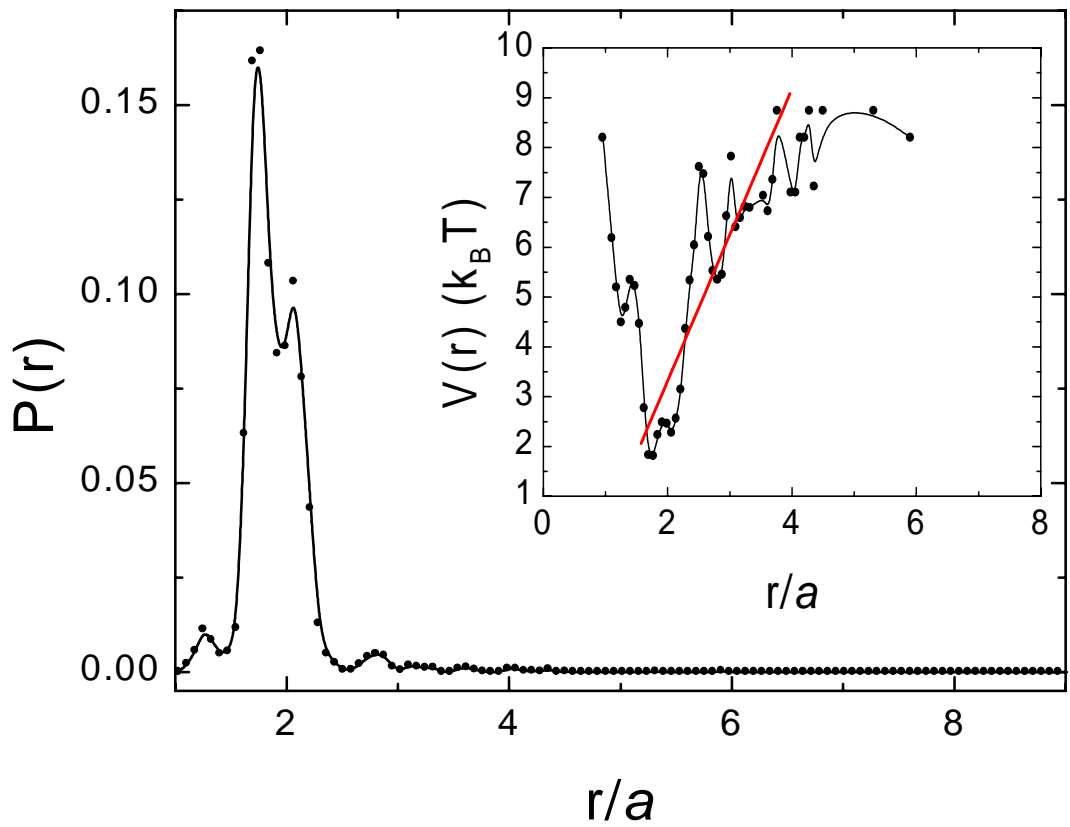
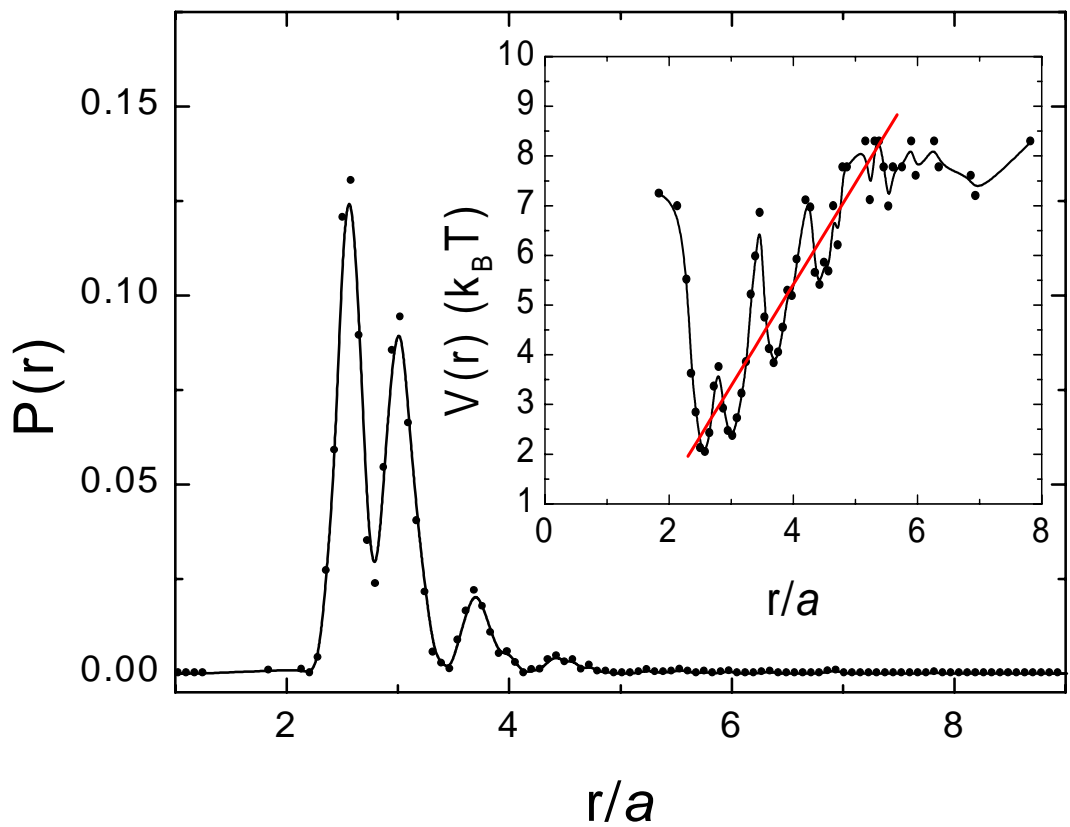
\begin{figure}
\caption{Pair interaction between dislocations. di(top) and mono-vacancies(bottom). $P(r)$ was determined from a few thousand snap-shots of the system by measuring the distribution of the separation of the two 5-fold coordinated particles found. $V(r)/k_{B}T$ was estimated from $-logP(r)$. The solid curves are spline fits and the straight lines are linear fits to the particular regions of data.}
\label{P(r)}
\end{figure}

Focusing on the case in which the point defects appear as a dislocation pair and assuming an interaction $V(\vec{r})$ between the two dislocations, we would expect Boltzmann statistics for the probability of observing the two dislocations a certain distance apart, namely: $P(\vec{r})\propto exp(-V(\vec{r})/k_{B}T)$. In addition, particle conservation dictates that dislocations can only glide parallel to their Burgers vector, so $\vec{r}=\vec{r}_{c}+\vec{r}_{g}$, with $\vec{r}_{c}\approx3\vec{a}(2\vec{a})$ for di(mono)vacancy and $\vec{r}_{g}\parallel \vec{b}$. We measured $P(r)$, $r=|\vec{r}|$, for the mono and di-vacancies, using a few thousand snapshots of the system. The separation $r$ of the two dislocations is identified as the separation between the two five-fold coordinated particles where the extra rows of particles terminate. Our results (Fig.\ref{P(r)}) show a rather rapid decrease in $P(r)$ as $r$ goes beyond a couple of lattice constants, together with a modulation, more clearly seen in the case of the di-vacancy. Since $V(r)\propto-logP(r)$, the interaction of the two dislocations turns out to increase with $r$, with an average slope $2.9\pm0.4\text{ and }2.0\pm0.2  k_{B}T/a$ for the case of mono and di-vacancy respectively. The modulation corresponds to a Peierls-like energy barrier\cite{disloc} that comes from the discreteness of the lattice. A final point is that the core region of dislocations has a typical size of a few $a$. Therefore, our measurements of $V(r)$ are in a regime where the cores of the two dislocations overlap and linear elasticity theory is not applicable.

%\section{Conclusions}
In summary, we demonstrate a novel way of introducing point defects in colloidal crystals through manipulation of individual particles with optical tweezers. Using digital video microscopy, we identified the topological configurations of the particles around the defect and showed how they are connected with the energetic configurations found by numerical work in \cite{defect_configs}. Topological analysis of the possible configurations reveals that the point defects can appear as dislocation $\it{multipoles}$ with zero total Burgers vector. The individual dislocations are bound by an attractive interaction. Di-vacancies in particular are observed to dissociate into two dislocations that can separate several lattice spacings before recombination.

%\acknowledgments
We acknowledge helpful discussions with Prof. S.C. Ying.
This work was supported by NSF (grant DMR-9804083), the Petroleum Research Fund, and the Research Corporation. XSL acknowledges the support of the A.P. Sloan Foundation.

% now the references. delete or change fake bibitem. delete next three
%   lines and directly read in your .bbl file if you use bibtex.

% figures follow here
%
% Here is an example of the general form of a figure:
% Fill in the caption in the braces of the \caption{} command. Put the label
% that you will use with \ref{} command in the braces of the \label{} command.
%
% \begin{figure}
% \caption{}
% \label{}
% \end{figure}

% tables follow here
%
% Here is an example of the general form of a table:
% Fill in the caption in the braces of the \caption{} command. Put the label
% that you will use with \ref{} command in the braces of the \label{} command.
% Insert the column specifiers (l, r, c, d, etc.) in the empty braces of the
% \begin{tabular}{} command.
%
% \begin{table}
% \caption{}
% \label{}
% \begin{tabular}{}
% \end{tabular}
% \end{table}

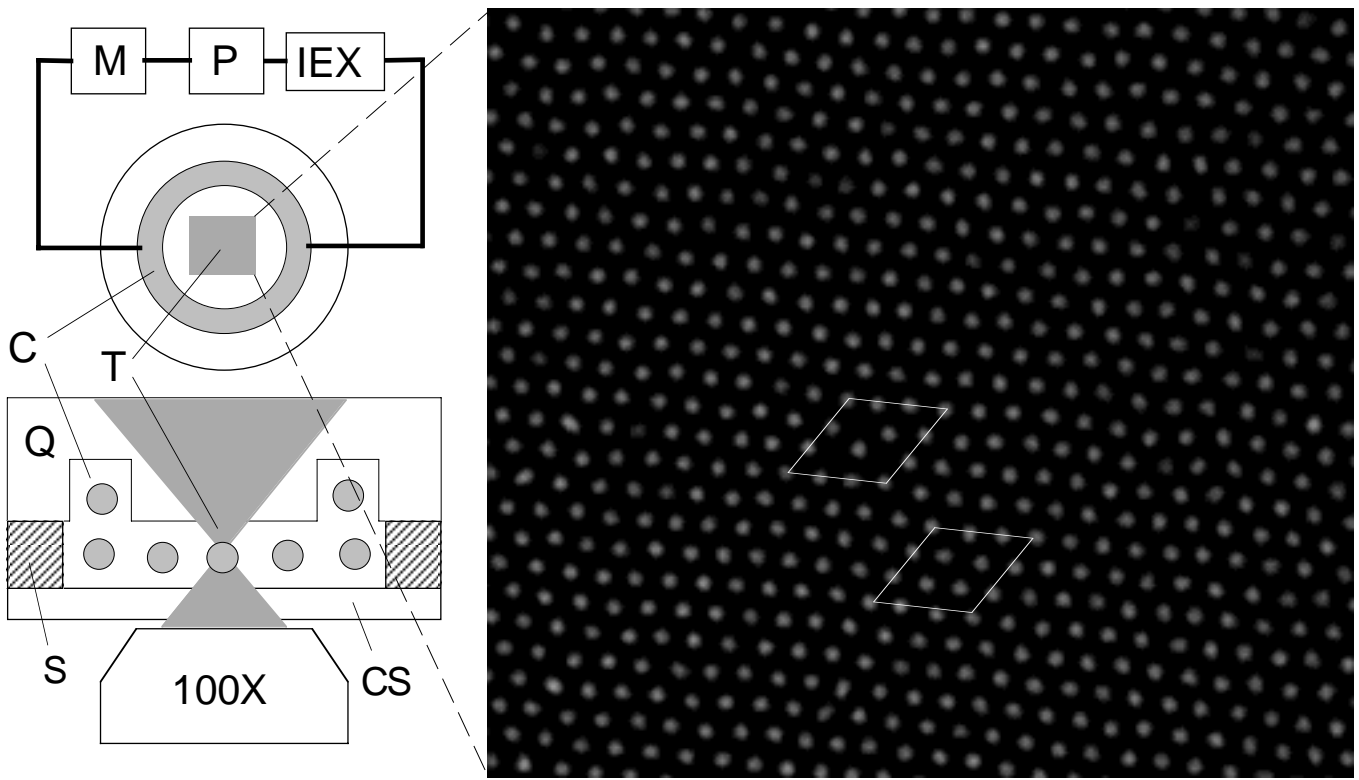
\begin{figure}
\caption{\textit{(Left)}Experimental setup (not to scale): The sample cell consists of a 1/2" diameter quartz disk(Q) and a quartz coverslip(CS) glued together. The distance between the two surfaces was controlled by a patterned thin ($\approx2\mu m$) polymer film (S) (Dow Chemical Co., CYCLOTENE), which served both as a spacer and as an adhesive. The cell was connected to an external circulation circuit containing a few mL of suspension, driven by a peristaltic pump (P) (VWR) and including a conductivity meter (M) (VWR, model 1054, flow-through cell) and a column with mixed-bed ion-exchange resin (IEX) (BIO-RAD, AG 501-X8(D)). (C) Indicates a 0.5mm circulation channel in contact with the two-dimensional region and (T) a particle trapped with the optical tweezers. \textit{(Right)}Micrograph of an isolated di-vacancy in a two-dimensional colloidal crystal. The $4\times4$ diamond in the enlarged central region contains 4 particles in the perfect lattice but only 2 when it encloses the core of the di-vacancy.}
\label{exp_setup}
\end{figure}

\begin{figure}
\caption{Configurations of mono-vacancy: (\textbf{A})split $(SV)$, (\textbf{B}) symmetric $(V_{3})$, (\textbf{C,D}) crushed $(V_{2a},V_{2b})$. Nomenclature adopted from Ref.15. Top left insets in every figure show configuration of the vectors $\vec{n}_{ij}$ defined in the text. Top right insets show the arrangement of miscoordinated particles in the core of the defect. The relative free energies of the different configurations were estimated to be (in $k_{B}T$): $F_{SV}=0.86, F_{V3}=1.10, F_{V2a} + F_{V2b}=1.41$.
Configurations of di-vacancy: (\textbf{E,F})split $(SD_{e,b})$, (\textbf{C}) crushed $(D_{2})$, (\textbf{D}) symmetric $(D_{3})$. Free energies in $k_{B}T$ are: $F_{D2}=0.7, F_{SD}=1.1, F_{D3}=2.1$.}
\label{pic_configs}
\end{figure}

\begin{references}
\bibitem{c1} The two-dimensional region extended over a few hundred lattice constants. We used optical tweezers to drag impurities, such as aggregates of particles, away from a central region 40$\times$40 lattice constants, where the experiments were performed. In these experimental conditions (density of particles and screening length) the resulting two-dimensional crystal was deep in the solid phase, with no thermally generated defects.
\bibitem{di_creation} The same procedure often produced a di-vacancy instead of a mono-vacancy. We are not able to observe the creation of the defect in detail because the entire process is completed within a few frames and the reflection of the laser beam obscures imaging around the tracked particle. The evidence available suggests that dragging a particle away from its lattice point and parallel to a crystal axis can push an entire row of particles. The tweezed particle is moved to a new lattice site leaving a first vacancy behind. The row buckles several $a$ away, creating an interstitial. Continuing to drag the particle further creates a second vacant site, resulting in a di-vacancy.
\end{references}
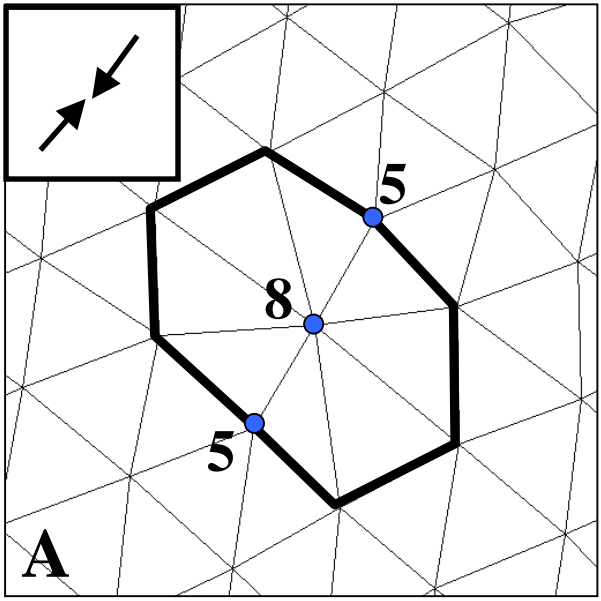
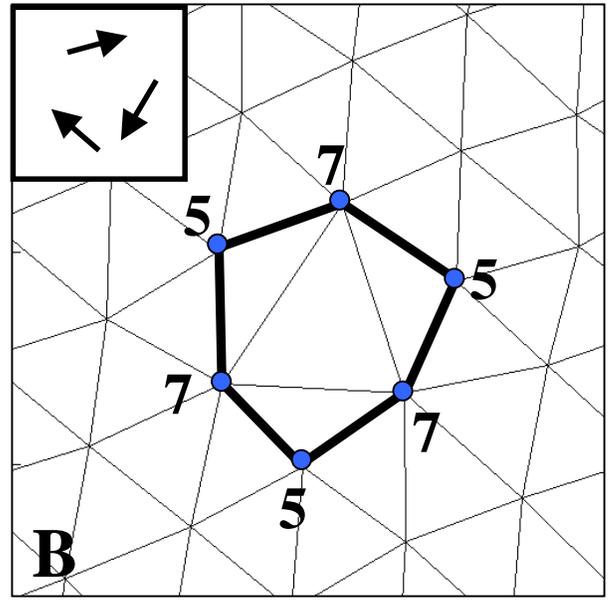
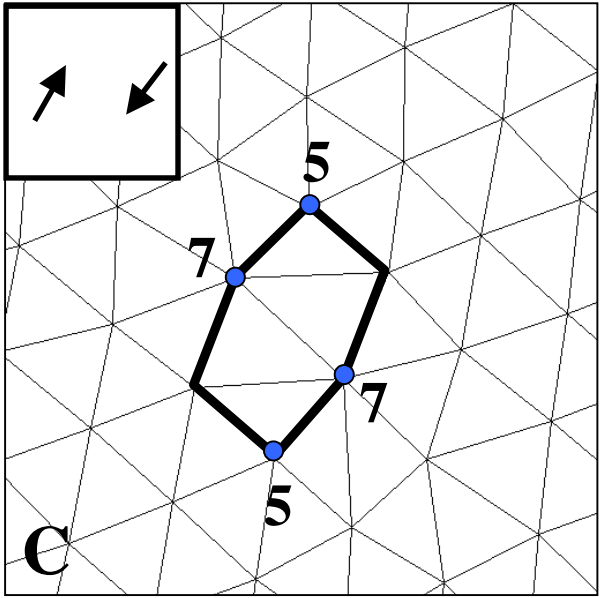
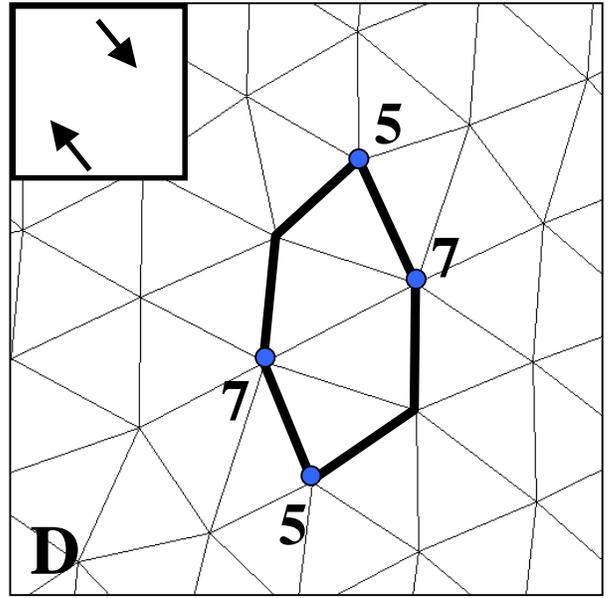
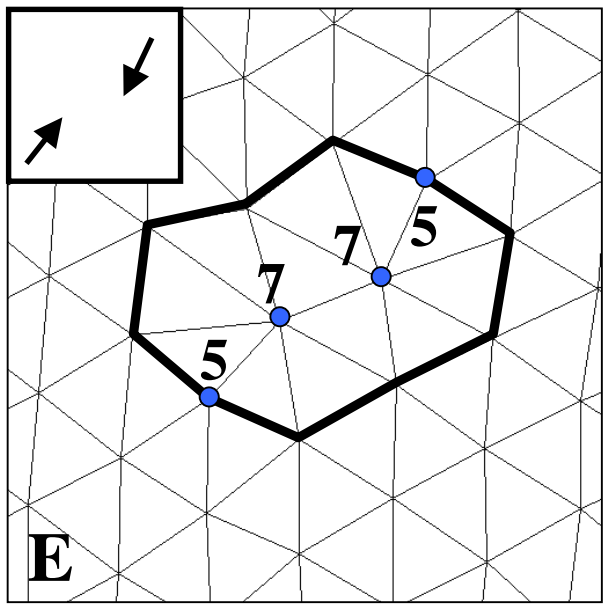
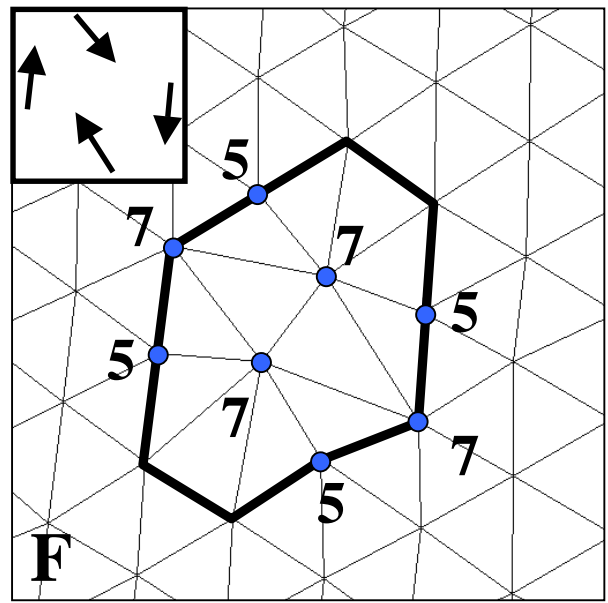
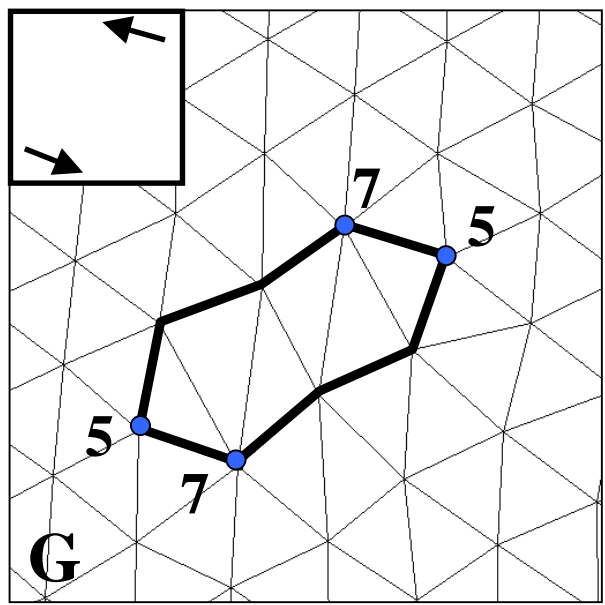
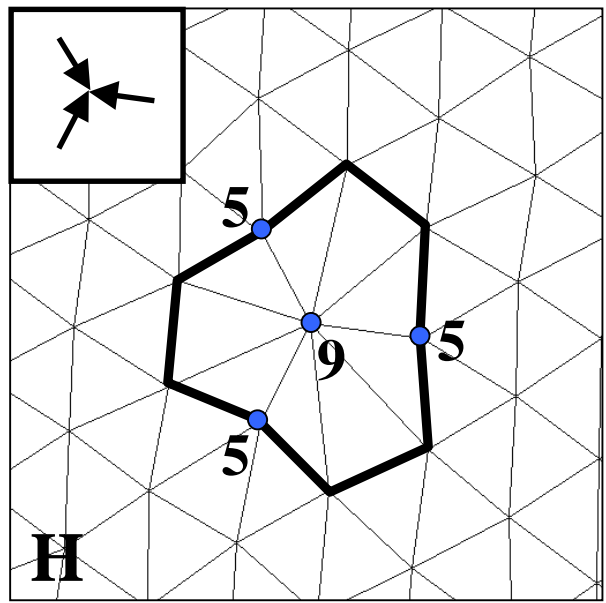
\end{document}